# On optimization of the metal ion production by Electron Cyclotron Resonance Ion Sources

V. Mironov, S. Bogomolov, A. Bondarchenko, A. Efremov, K. Kuzmenkov, V. Loginov, D. Pugachev

*Joint Institute for Nuclear Research, Flerov Laboratory of Nuclear Reactions,
Dubna, Moscow Reg. 141980, Russia*
E-mail: vemironov@jinr.ru

Abstract: The three-dimensional NAM-ECRIS model is applied for studying the metal ion production in the DECRIS-PM Electron Cyclotron Resonance Ion Source. Experimentally measured extracted ion currents are accurately reproduced with the model. Parameters of the injection of metal vapors into the source are optimized. It is found that the axial injection of the highly directional fluxes allows increasing the extracted ion currents of the highly charged calcium ions by factor of 1.5. The reason for the gain in the currents is formation of internal barrier for the ions inside the plasma, which increase the ion extraction and production efficiency. Benefits of injecting the singly-charged calcium ions instead of atoms are discussed.

Intense and stable beams of the metal ions are requested in a variety of applications. In particular, accelerated $^{48}$Ca ions are routinely used in experiments on synthesis of the super-heavy elements [1] at the FLNR accelerator complexes. The Electron Cyclotron Resonance Ion Sources [2] are considered to be the best choice for production of the medium and high charged ion beams, being able to deliver ions in the mass range from hydrogen to uranium, with ion currents as large as ~1 mA for the relatively light elements ($Ar^{8+}$-$O^{6+}$) and ~0.1 mA for the heavy and highly charged $U^{35+}$ ions. The sources are open magnetic traps with plasma heated by absorption of microwaves at the electron cyclotron resonance. The source magnetic field is formed either by a set of axially symmetric solenoids or by permanent ring magnets, combined with the multipole magnetic field to ensure that the magnetic field is increasing from the trap center in any direction. Electron component of the ECRIS plasma is confined by reflections of electrons from the regions with increasing magnetic field; ions are produced by sequential ionizing electron-ion collisions and reach the high charge state if the ion confinement time and electron density are high enough.

For production of the metal ions, corresponding vapors should be injected into the source chamber. Resistively heated micro-ovens (effusion cells) can be used for those elements that can be evaporated at the relatively low temperatures in the range below 1000 °C. For the refractory elements, there are used either high-temperature ovens or injection of the metal-containing volatile organic vapors [3]. Sputtering of metal rods with plasma ions is tested to be convenient for such elements as vanadium [4] and uranium [5]. Production of the ions in the desired charge state should be done with high efficiency: typically, experiments with the accelerated ion beams last continuously for a few days or months. Intervals between the oven reloads with material should be as long as possible. Some isotopically enriched materials are expensive and their consumption should be minimized. Therefore, geometry of the metal vapor injection has to be optimized for the efficient and stable source operation.

We numerically investigate the calcium ion production in DECRIS-PM ion source [6] with making an accent on selection of the optimized conditions for the vapor injection from micro-oven. The Numerical Advanced Model of ECRIS (NAM-ECRIS) is used [7] for the calculations. The paper is organized in the following way: first, details of the DECRIS-PM are given. The numerical model specific features are discussed next. Results of calculations are presented for the source in the configuration that is currently in use. Then, we discuss the possible ways for optimization of the source performance by varying the micro-oven position and angular distribution of the injected calcium fluxes. Benefits of injecting the singly-charged calcium ions instead of atoms are discussed in the end.

## *Description of the DECRIS-PM source.*

The magnetic fields of the source are formed by a set of permanent magnets. The solenoidal component is produced by the ring-shaped magnets with the radial and axial magnetizations; the hexapole component is produced by the 24-segment Halbach magnet. The magnetic field on the axis is 1.34 T at the injection flange, 1.1 T at the extraction electrode, and the minimal magnetic field is 0.42 T at the source center. The hexapole field at the source chamber radial walls is 1.1 T. The source chamber dimensions are 23 cm in length and 7 cm in diameter. The ions are extracted through extraction aperture of 1-cm in diameter at the extraction side of the source. At the injection flange (Fig.1), there are installed the waveguide for injection of 14.5 GHz microwaves, the aluminum biased electrode with the diameter of 3.0 cm, the gas injection port and the port for the micro-oven installation. Both ports are located off-axis at the radius of 2.3 cm. The flange is perforated for better pumping of the source chamber, the holes are of 0.6 cm in diameter and cover 25% of the surface not shielded by the biased electrode. The plasma shape close to the injection flange is a triangular star, with the densest plasma at the source axis. Plasma sputter pattern on the biased electrode is seen in Fig.1, as well as the plasma star orientation in respect to the injection ports.

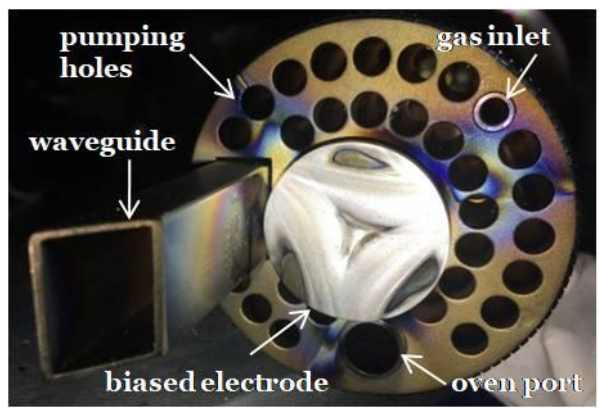

Fig.1. Injection flange of DECRIS-PM.

The micro-oven is placed inside the oven port. The oven includes the resistively heated filament, the thermal shield and the crucible filled with a solid material. The inner diameter of the crucible is 2 mm. For such working materials as Li or Bi, exit part of the crucible is closed with a wire plug to prevent a spilling of occasionally melted material into the source chamber. The wire plug makes the angular distribution of the ejected vapors close to isotropic. For calcium the wire plug is usually not used, but the crucible is filled with calcium such that the atomic flux collimation by the exit aperture can be neglected.

For injection of calcium atoms, the operational crucible temperature is around 400-600 °C. Variations in the temperature allow regulation of the atom flux; typical material consumption is around of 1.5 mg/h for the optimized extracted currents of $Ca^{10+}$ ions close to 150 μA. For the experiments on the super-heavy element synthesis, the isotopically enriched material is used with the $^{48}Ca$ isotope abundance of

67%; during the source tests at the test-bench, the natural calcium is used with the $^{40}$Ca abundance of 97%.

To prevent the vapor condensation on the source chamber walls, thin tantalum screen (0.2 mm foil) is inserted into the chamber such as to cover the chamber radial walls and the plasma electrode at the extraction side, with the axial hole in the screen for the ion extraction. The injection side of the source is left un-shielded. The thermal contact of the screen with the water-cooled chamber is bad, and the screen is heated by the injected microwaves up to ~ (300-550) °C according to measurements [8].

In most cases, we use helium as the support gas. Other support gases such as oxygen and nitrogen were tested, but were found less effective for the calcium ion production. Extracted currents of He$^{1+}$ ions are around (600-800) µA in the optimized conditions of the source. The injected microwave power is (250-500) W, the biased electrode negative voltage is (100-500) V.

### *Numerical model.*

For calculations of ECRIS performance in a mix of helium and calcium ions, we use the NAM-ECRIS numerical model [7] with the source parameters described in the previous section. The model is a set of iteratively running 3D particle-in-cell modules that simulate the electron and ion dynamics in the plasma. The model uses the COMSOL Multiphysics® software for simulations of the microwave interaction with the magnetized ECR plasma in the source chamber. The COMSOL's model produces the spatial and phase distributions of the microwave electric field amplitude. The distributions are used by the electron module to calculate the electron heating by the microwaves. In the module, electron scattering in collisions with the plasma electron and ion components is considered by using the corresponding spatial distributions from the previous runs of the electron and ion modules. The electron module produces the energy and spatial distributions of electron component, as well as the globally defined electron life time. The ion module uses this information to trace the ion dynamics in the plasma.

Calculations begin with assuming a seed electron spatial distribution and are continued in the iterative way until the converged solution is obtained. During the iterations, plasma is characterized by the electron density at the plasma halo, which value is fixed. The converged solution provides the main plasma parameters such as the plasma density and the electron energy distribution functions, the fluxes of the injected material that are needed for reaching the plasma density with the selected value, and the extracted ion currents. Also, plasma potential distributions are obtained in the model, as well as the electron and ion life times. It is possible to accurately study the source responses to variations of source parameters in the essentially self-consistent way.

The ion module of the NAM-ECRIS uses a fixed number of computational particles, 4×10$^5$ in these specific runs. Relative fluxes of helium and calcium atoms into the source are controlled by allocating different portions of the total particle number for He and Ca particles. Whenever a particle leaves the source chamber, it is returned back into the computational domain through the injection port. Helium atoms are injected with energies selected from the Maxwell-Boltzmann distribution for the room temperature. Angular distribution of the injected helium atom velocities is always isotropic. Full reflection of helium from the source chamber walls is assumed, except of those particles that are lost at the extraction aperture or at the regions outside the biased electrode with a probability of 25% corresponding to the total transparency of the pumping channels. At this, we neglect the details of the channel spatial distribution.

Both helium and calcium ions are assumed to be accelerated close to the chamber walls normally to the surface up to 25×Q eV, where Q is the corresponding ion charge state and 25 V is the estimation of the plasma potential drop in the sheath. Neutralized particles are then isotropically reflected from the surface retaining some fraction of their primary kinetic energy. For helium, we use the thermal accommodation coefficient from [9] for the Ta surface (temperature of the surface is set to 300°C), for calcium we assume that the coefficient is 5%, same as the experimental data for argon hitting the stainless-steel surfaces [10]. We omit all details concerning the calcium atoms' interaction with the walls, which are difficult to treat accurately. Generally, calcium condensation at the cold surfaces is always seen when inspecting the source after prolonged operation, especially pronounced at the injection flange. Screen surface is not noticeably covered by the deposited calcium. The deposited calcium atoms are either re-evaporated from the surfaces thermally, if the local temperature is high enough, or are sputtered by impacts of ions and energetic atoms. Numerically, we introduce the sticking probability for calcium hitting the walls, using this value as the free parameter in the model. If the calcium particles are lost from the system at the walls, through the extraction aperture and through the pumping channels, they are returned back at the oven port position with the velocity sampled from the Maxwell-Boltzmann distribution with the temperature of 600 °C. In the default conditions, the angular distribution of calcium atom is isotropic, the same as for the helium atoms. The distribution can be generated more peaked along the injection axis by changing the ratio between the perpendicular and the longitudinal temperatures of the injected atoms, with keeping the total energy of the particles the same.

Ionization rates for helium and calcium are calculated from the electron energy distributions obtained in the electron module. The Burgess-Chidichimo [11] cross-sections are used. The excitation-autoionization contributions to the ionization rates are obtained from the data of [12], available only for the Maxwell-Boltzmann distributions of electron energies. The authors of [12] claim that their rates can differ from the experimental data by factor of 2 or higher; we found that it is necessary to double the excitation-autoionization rate for $Ca^{9+}$ ions to reach the better correspondence between the calculated and experimentally measured charge-state distributions of the extracted calcium ions. The charge-exchange collisions of ions with helium and calcium atoms are calculated by using the Langevin rates. The ion heating due to electron-ion collisions and ion scattering in ion-ion collisions are traced as described in [7] using the Monte-Carlo collision techniques. Calculations are done for 500 W of the injected microwave power and for the reference electron density of $1.25 \times 10^{11}$ cm$^{-3}$. The $^{40}$Ca isotope is considered as the working material.

### *Results of the calculations.*

We begin with presentation of the main plasma parameters obtained with the off-axis injection of calcium atoms and with isotropic angular velocity distribution of the injected atoms. The wall-sticking coefficient is set to zero, and the calcium atoms are lost from the system only at the extraction aperture and through the pumping channels at the injection side of the source. The helium-calcium mixing ratio is kept fixed as (0.9-0.1) throughout the calculations.

The electron density spatial distributions are shown in Fig.2 for the profiles along the source axis (A) and in the transversal direction at the source center, z=11.5 cm (B).

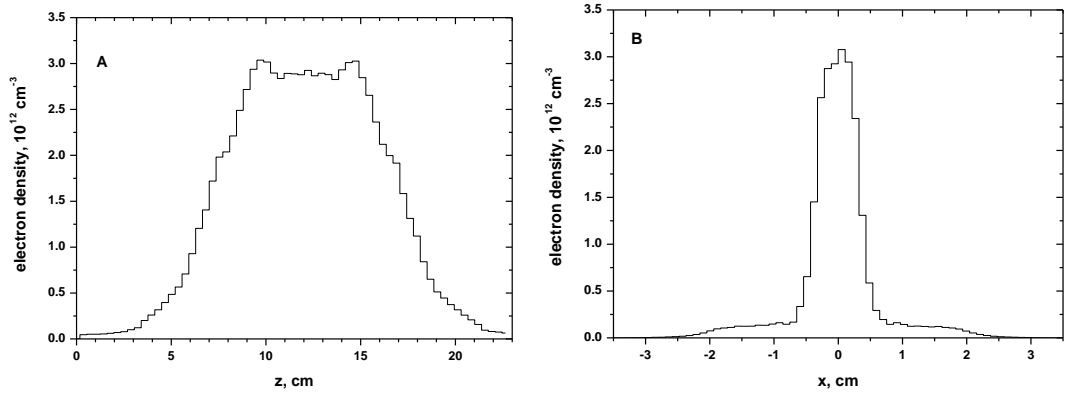

Fig.2. Electron density profiles along the source axis (A) and in the transversal direction (B).

The plasma is strongly localized at the source axis due to trapping of electrons by retarding voltages at the extraction aperture and at the biased electrode [7]. Longitudinally, plasma is preferentially confined inside the ECR volume. The maximal electron density is at the level of $3 \times 10^{12}$ cm$^{-3}$, slightly above the cut-off value for the 14.5 GHz microwaves. The electron energies are calculated separately for the particles in the plasma core (for the distances from the axis less than 5 mm) and in the halo, as well as inside and outside the ECR volume. The distributions are fitted by a sum of two decreasing exponents representing the warm and hot components of the electron population. Energies of electrons inside the core and in the ECR zone are best characterized as a sum of 30 and 77 keV decaying exponents with the relative contribution of warm electrons equal to 72%. For the electrons inside the core and outside the zone, the distribution is 3.5 and 110 keV, with the warm population of 22%. The corresponding parameters for the halo electrons inside the zone are 20 and 60 keV with the warm electron contribution of 55%, and 77 keV electrons outside the zone, with no warm electron contribution. The electrons outside the ECR zone are relatively hot because they can interact there with microwaves due to Doppler and relativistic shifts of the cyclotron resonance. The globally defined electron life time is 0.28 ms in the given conditions.

The plasma potential is calculated without considering the potential drop in the sheath, which is estimated to be around +25 V.

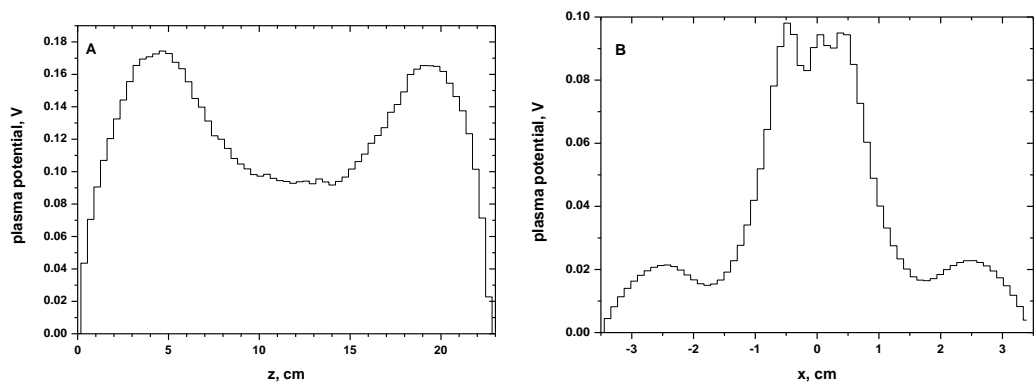

Fig.3. Longitudinal (A) and transversal (B) profiles of the plasma potential.

The potential longitudinal and transversal profiles are shown in Fig.3. The largest potential is at the level of 0.17 V, with the potential in the plasma center of ~0.09 V. The plasma localization on the source axis results in the narrow potential peak there. The potential dip is developed inside the dense parts of the plasma in the ECR volume, which value is defined by the ion temperatures and by the need to equilibrate the electron and ion losses. The dip value is 0.08 V in the given conditions. It has been shown in [7] that the transversal electric fields associated with the plasma potential gradients strongly affect the ion diffusion toward the radial walls of the source and the extracted ion currents.

The calculated charge-state distribution of the extracted calcium ions is shown in Fig.4 as the red columns. The experimentally measured currents are shown as open circles. The spectrum is measured at the source test-bench with injection of $^{40}$Ca in the mix with helium; currents of $Ca^{10+}$ cannot be determined due to overlap with $He^{1+}$ ions. The source parameters were tuned to maximize the $Ca^{11+}$ current.

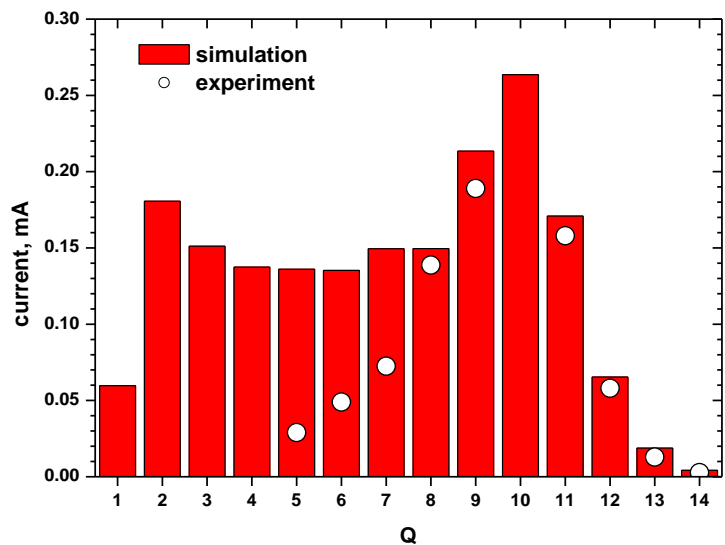

Fig.4. The simulated charge-state distribution of the extracted calcium ions (red columns) and the experimentally measured currents (open circles).

For the charge states above 8+, the model accurately reproduces both the shape of the charge-state distribution and the absolute values of the extracted ion currents. Strong deviations from the experimental values are seen for the relatively low charge states, probably due to large losses of these ions in the extraction gap and in the beam line. The calcium consumption is calculated by the model to be 1.3 mg/h, being compared to the estimated experimental value of ~1.5 mg/h. The ion temperatures are around 0.2 eV, with the neutral calcium temperature of 0.07 eV. For the given gas-mixing ratio, the helium ion currents are calculated to be 0.77 mA and 0.38 mA for $He^{1+}$ and $He^{2+}$ ions respectively.

The total efficiency of the calcium ion extraction is 45%, with the rest particles being lost at the injection flange through the pumping channels, plus small contribution of neutral calcium going through the extraction aperture. Contribution of $Ca^{10+}$ ions into the total extracted particle flow through the extraction aperture is 6.7%, which gives the extraction efficiency for these ions of 3% for the selected zero sticking coefficient of calcium on the walls. Incorporation of the sticking into the model results in increased material consumption and in decrease of the extraction efficiency. If we set the sticking coefficient to 1.0 (full absorption of calcium atoms after hitting the walls), then the calcium flow into the chamber should be increased to as much as 65 mg/h to maintain the plasma density at the requested level for the mixing ratio of helium/calcium of (9:1), with the corresponding drop in the extraction efficiency. By comparing the calculated and experimental values, we conclude that the sticking of calcium on the walls is small in the given conditions of the source. Boost in the extraction efficiency is observed if the particle losses through the pumping channels are blocked. In these conditions, the calcium consumption becomes 0.65 mg/h, decreased by factor of 2 compared to the default value. The extracted ion currents are not sufficiently affected by blocking the injection flange losses: the total extraction efficiency is calculated as 95%, and the $Ca^{10+}$ extraction efficiency is 6.8%.

The extraction efficiency of ions in ECRIS is affected by many other factors apart from the particle escape from the source chamber. In particular, faster ionization rates can increase the extraction efficiency for the highest charge states, as well as the mixing ratio between the support gas and calcium, which influences the plasma potential dip and the ion life times; charge-exchange collisions between the calcium ions and calcium and helium atoms decrease the mean charge state of ions inside the plasma and the ion extraction efficiencies. Also, not all produced ions with the desired charge state are going through the extraction aperture, but are lost elsewhere. In the conditions presented in Fig.4, the ratio between the total flux of $Ca^{10+}$ ions to the chamber walls and the flux into the extraction aperture (the transport efficiency) is 0.45. The radial losses of the $Ca^{10+}$ ions are relatively small due to strong localization of the ions at the source axis, but the fluxes toward the extraction and injection sides of the source are approximately the same. It is worth to investigate whether it is possible to redirect more calcium ions toward the extraction by optimizing the metal injection conditions.

We numerically study the benefits of the on-axis localization of the metal injection port. The results of calculations are shown in Fig.5 for different angular distributions of the injected fluxes. We use the same plasma parameters (plasma density and electron/ion life time) as for the off-axis injection. Helium atoms are injected off-axis.

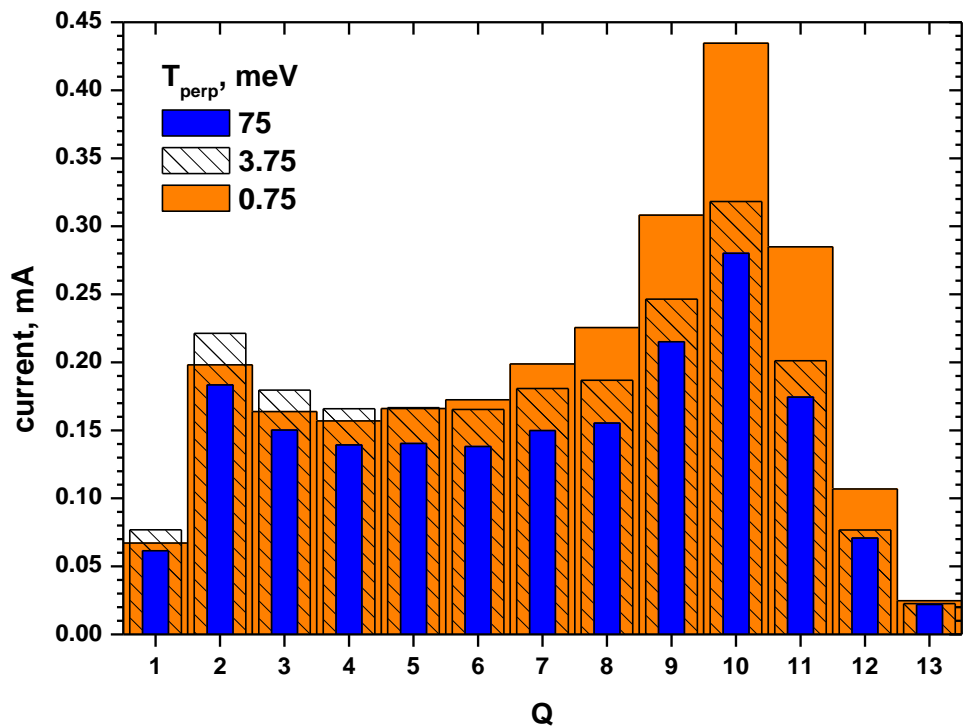

Fig.5. Charge-state distributions of the extracted calcium ions for the on-axis orientation of the metal injection port and for different angular distributions of the injected flux. The perpendicular temperatures of injected calcium atoms are 75 meV (isotropic value) for the blue columns, 3.75 meV for the dashed columns, and 0. 75 meV for the orange columns.

For the isotropic injection of atoms, the extracted ion currents remain the same as for the case of the off-axis injection. The metal consumption is the same 1.3 mg/h, and the extraction efficiencies are not changed. The more directional injection results in the increase of the currents for the relatively highly charged ions. For the $Ca^{10+}$ ions, current is increased up to 0.43 mA compared to 0.29 mA for the isotropic injection for the perpendicular temperatures of 0.75 meV; the longitudinal temperature is set to 223.5 meV in this case to keep the total mean energy of the injected atoms constant.

The calcium consumption rate remains unchanged with injection of the directional fluxes, but the extraction efficiency of the $Ca^{10+}$ ions (4.7%) is increased by the factor of 1.6 compared to the isotropic injection. The transport efficiency is 54%, higher than for the default injection conditions by the factor of 1.2. Content of the highly charged $Ca^{10+}$ ions inside the ECRIS plasma is increased by ~15%, and the ion life time for these ions is increased from 0.59 ms to 0.61 ms. We calculate the ion life time for the given ion charge state as the interval between creation of the ion and the moment of its loss on the source walls or through the extraction aperture.

The reason for the extracted ion boosts with the directed injection of the atom fluxes can be understood by investigating the spatial distributions of calcium particles inside the source. In Fig.6, slices of the particle trajectories are shown in the transversal plane along the source axis. The brighter is the pixel color, the more particles pass through this point during their movement inside the source chamber. The color scale is shown in the top parts of the pictures. The left column of the Figure represents the trajectories of $Ca^{0+}$, $Ca^{1+}$ and $Ca^{10+}$ for the isotropic injection of calcium, the right column shows the corresponding trajectories for the highly directed injected fluxes. Neutral and singly charged calcium particles do not penetrate into the dense central parts of the plasma because of ionization into the higher charge states.

| Isotropic injection, $T_{perp}$=75 meV | Directed injection, $T_{perp}$=0.75 meV |
|---|---|
| $Ca^{0+}$ | $Ca^{0+}$ |

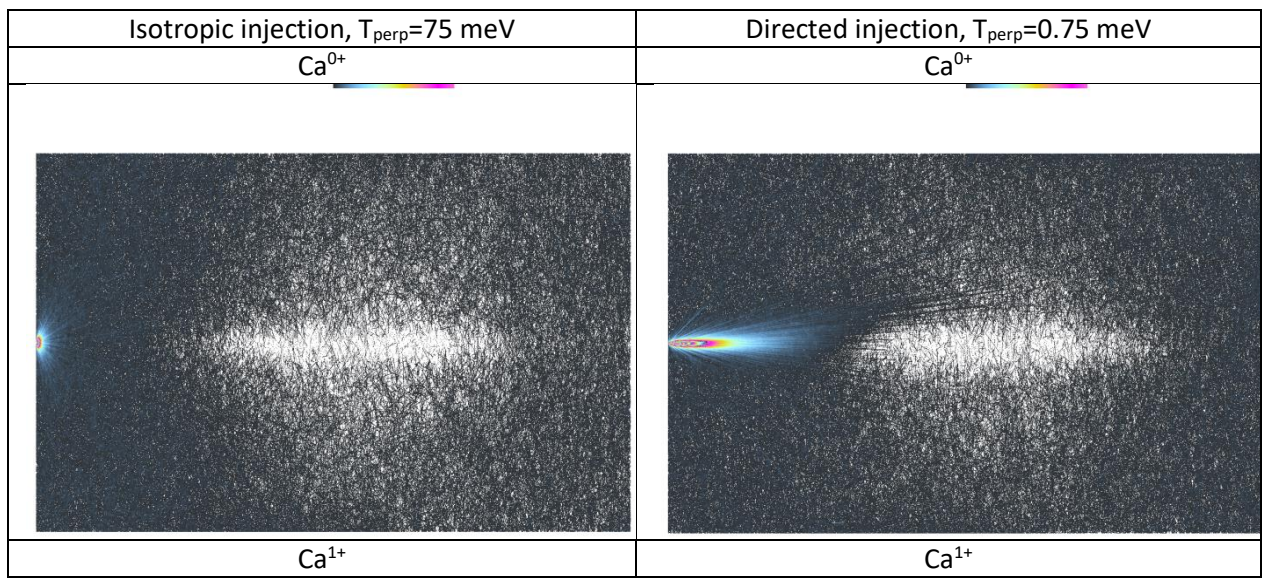

| $Ca^{1+}$ | $Ca^{1+}$ |

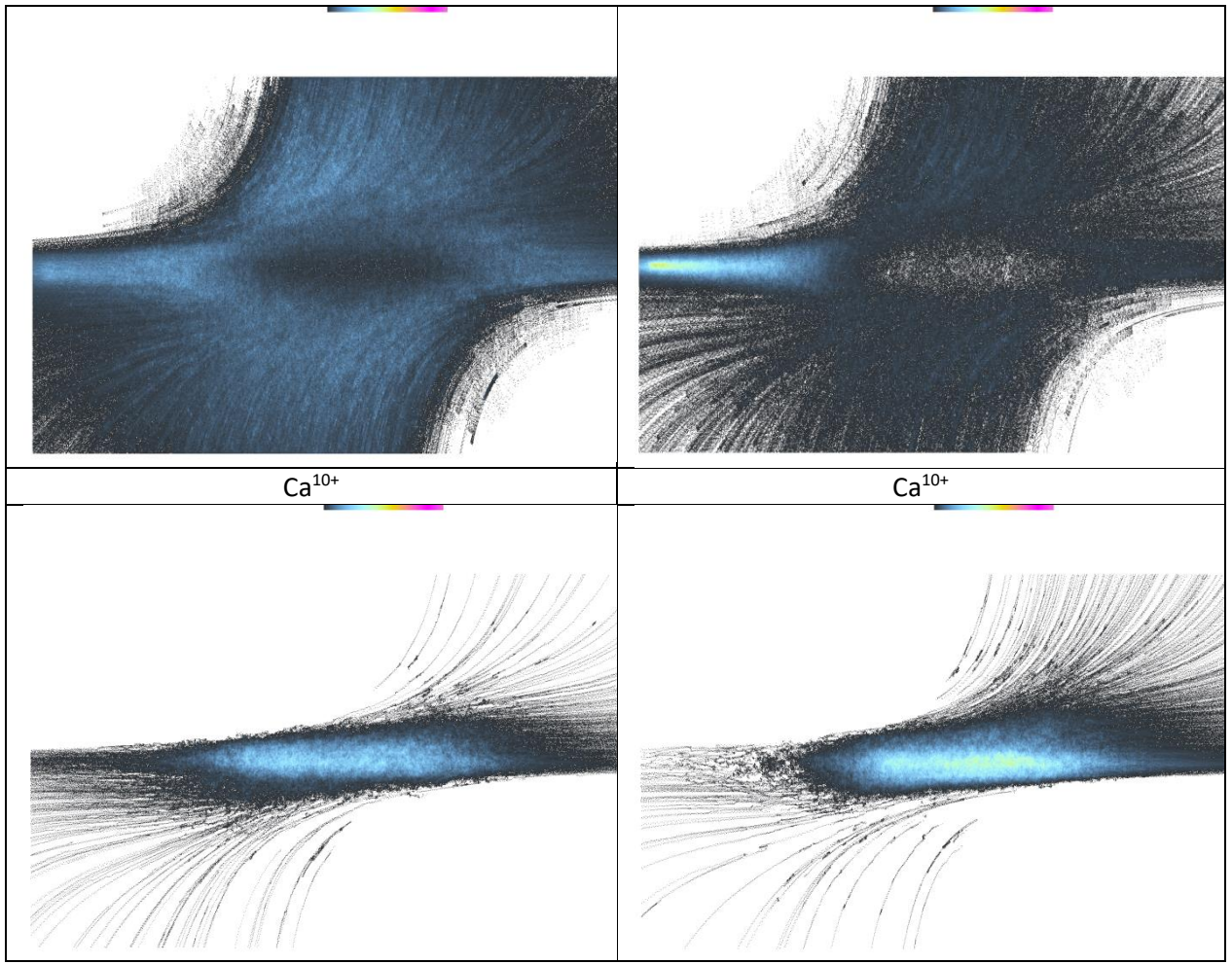

Fig.6. Trajectories of $Ca^{0+}$, $Ca^{1+}$ and $Ca^{10+}$ in the transversal slice of the source chamber for the isotropic injection (left) and for the directed injection (right).

For the directed injection, much more singly charged calcium ions are created at the injection side of the source along the axis compared to the isotropic injection. This results in formation of the internal transport barrier for the highly charged calcium ions, which fluxes become to be more directed toward the extraction aperture due to the "injection wind". The increased transport efficiency to the extraction is the result of the directed injection, as well as the better confinement of the highly charged ions.

The internal barrier can be formed more efficiently if injecting the singly charged calcium ions instead of the neutrals, with the larger energies of the injected particles and their higher directionality. We investigate this scheme by calculating the extracted ion currents with $Ca^{1+}$ injection along the source axis with the following initial conditions of the injection: the beam size is 5 mm in diameter, ion velocities are directed along the z-axis, ion energies are varied from 0.25 to 5 eV (the energies are defined at the injection plane after ion retardation by the positive plasma potential of ~25 V). The charge-state distributions of the extracted calcium ions are shown in Fig.7 for different injection energies of 025, 1 and 5 eV.

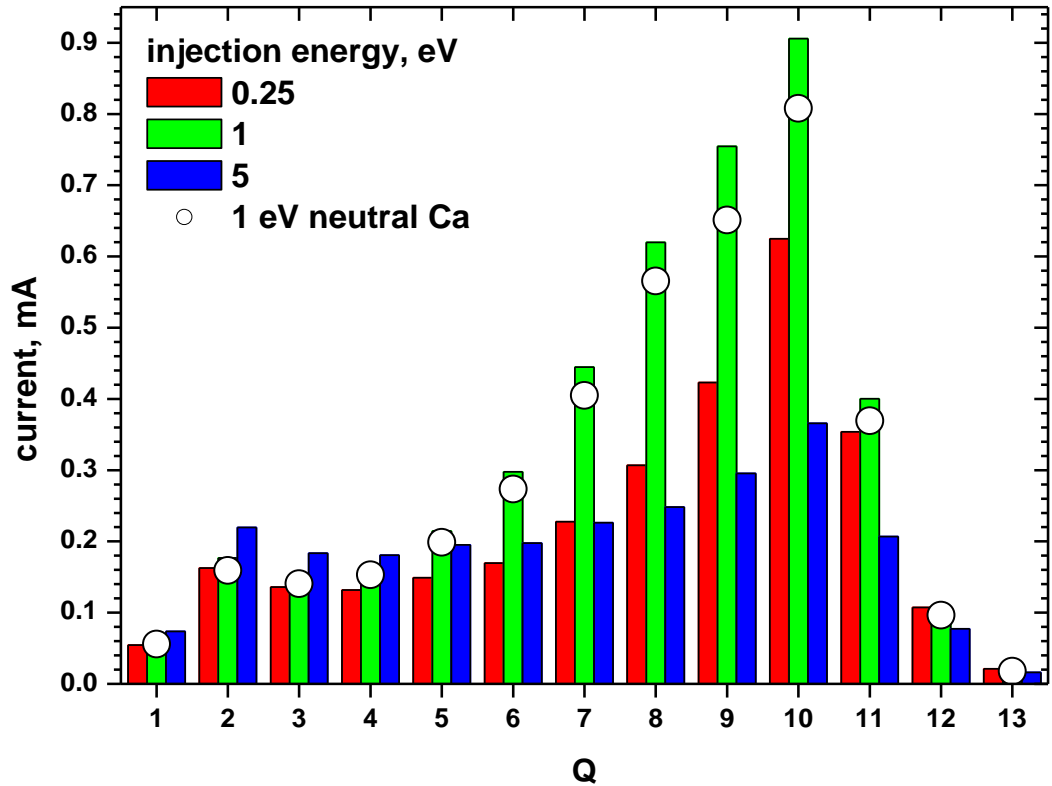

Fig.7. Charge state distributions of the extracted calcium ions at injection of the singly charged calcium ions with energies of 0.25 (red), 1 (green) and 5 eV (blue). Charge state distribution for injection of the neutral calcium with energy of 1 eV is shown as open circles.

There is the optimal 1-eV injection energy of the calcium ions that results in the increase of the extracted ions by approximately factor of 2 compared to the on-axis injection of the neutral calcium with relatively low energy of 75 meV. The ion transport efficiency is increased up to 72% at the optimized injection, while the calcium consumption is only slightly larger than in the default conditions, 1.6 mg/h, which corresponds to 1 mA of the injected calcium current. With the larger injection energy of calcium ions, efficiency of ion transport to extraction is decreasing to 52%, and current of $Ca^{10+}$ ions is ~0.3 mA, less than for the optimized injection energy. The benefits of the described injection scheme are not due to the charge state of the injected particles: injection of the neutral atoms with the same energies and angular velocity distribution gives, due to the fast ionization of the injected fluxes, the extracted ions currents close to the values obtained with injection of the singly charged calcium. The corresponding distribution is shown in Fig.7 as open circles. However, use of the singly charged injected ions allows optimizing the injection energy and the angular distributions. Technical problems of forming the ion beam with mA currents and energy of ~ (25-50) eV are beyond the scope of present calculations.

## *Acknowledgements*

This work was supported by the Russian Foundation for Basic Research under grant No. 20-52-53026/20.